\documentclass[11pt, a4paper]{article}
\pdfoutput=1
\usepackage{jcappub}

\usepackage{booktabs}
\usepackage{multirow}
\usepackage{amssymb}
\usepackage[export]{adjustbox}
\usepackage{floatrow}
\newfloatcommand{capbtabbox}{table}[][3.5in]
\newfloatcommand{capbfigbox}{figure}[][2.3in]
\usepackage{blindtext}
\usepackage{amsmath}
\usepackage{booktabs}
\usepackage{aas_macros}
\newcommand{\ie}{i.e.~}

\def\lsim{\mathrel{\raise.3ex\hbox{$<$\kern-.75em\lower1ex\hbox{$\sim$}}}}
\def\gsim{\mathrel{\raise.3ex\hbox{$>$\kern-.75em\lower1ex\hbox{$\sim$}}}}

\begin{document}

\hspace*{110mm}{\large \tt FERMILAB-PUB-16-609-A}

\vskip 0.2in

\title{Low Mass X-Ray Binaries in the Inner Galaxy: Implications for Millisecond Pulsars and the GeV Excess}

\author{Daryl Haggard,$^{a,b}$}
\emailAdd{daryl.haggard@mcgill.ca}\note{ORCID: http://orcid.org/0000-0001-6803-2138}
\author{Craig Heinke,$^c$}\note{ORCID: http://orcid.org/0000-0003-3944-6109}
\emailAdd{heinke@ualberta.ca}
\author{Dan Hooper$^{d,e,f}$}\note{ORCID: http://orcid.org/0000-0001-8837-4127}
\emailAdd{dhooper@fnal.gov}
\author{and Tim Linden$^g$}\note{ORCID: http://orcid.org/0000-0001-9888-0971}
\emailAdd{linden.70@osu.edu}

\affiliation[a]{McGill University, Department of Physics, 3600 rue University, Montreal, QC, H3A 2T8}
\affiliation[b]{McGill Space Institute, 3550 rue University, Montreal, QC, H3A 2A7}
\affiliation[c]{University of Alberta, Department of Physics, CCIS 4-183, Edmonton, AB, T6G 2E1}
\affiliation[d]{Fermi National Accelerator Laboratory, Center for Particle
Astrophysics, Batavia, IL 60510}
\affiliation[e]{University of Chicago, Department of Astronomy and Astrophysics, Chicago, IL 60637}
\affiliation[f]{University of Chicago, Kavli Institute for Cosmological Physics, Chicago, IL 60637}
\affiliation[g]{Ohio State University, Center for Cosmology and AstroParticle Physics (CCAPP), Columbus, OH  43210}

\abstract{If millisecond pulsars (MSPs) are responsible for the excess gamma-ray emission observed from the region surrounding the Galactic Center, the same region should also contain a large population of low-mass X-ray binaries (LMXBs). In this study, we compile and utilize a sizable catalog of LMXBs observed in the the Milky Way's globular cluster system and in the Inner Galaxy, as well as the gamma-ray emission observed from globular clusters, to estimate the flux of gamma rays predicted from MSPs in the Inner Galaxy. From this comparison, we conclude that only up to $\sim$\,4-23\% of the observed gamma-ray excess is likely to originate from MSPs. This result is consistent with, and more robust than, previous estimates which utilized smaller samples of both globular clusters and LMXBs. If MSPs had been responsible for the entirety of the observed excess, INTEGRAL should have detected $\sim$\,$10^3$ LMXBs from within a $10^{\circ}$ radius around the Galactic Center, whereas only 42 LMXBs (and 46 additional LMXB candidates) have been observed.}

\maketitle

\section{Introduction}

A statistically significant excess of GeV gamma rays has been observed from the region surrounding the Galactic Center~\cite{Goodenough:2009gk,Hooper:2010mq,Hooper:2011ti,Abazajian:2012pn,Gordon:2013vta,Hooper:2013rwa,Daylan:2014rsa,Calore:2014xka,TheFermi-LAT:2015kwa,Karwin:2016tsw}, with spectral and morphological characteristics consistent with the annihilation of dark matter particles. Although dark matter interpretations of this signal have generated a great deal of interest (see, for example, Refs.~\cite{Abdullah:2014lla,Ipek:2014gua,Izaguirre:2014vva,Agrawal:2014una,Berlin:2014tja,Alves:2014yha,Boehm:2014hva,Martin:2014sxa,Huang:2014cla,Cerdeno:2014cda,Okada:2013bna,Freese:2015ysa,Fonseca:2015rwa,Bertone:2015tza,Cline:2015qha,Berlin:2015wwa,Caron:2015wda,Cerdeno:2015ega,Liu:2014cma,Hooper:2014fda,Arcadi:2014lta,Cahill-Rowley:2014ora,Ko:2014loa,McDermott:2014rqa}), astrophysical origins for this emission have also been proposed. The leading astrophysical explanation for the observed excess is emission from a large population of unresolved millisecond pulsars (MSPs)~\cite{Hooper:2010mq,Hooper:2011ti,Abazajian:2012pn,Gordon:2013vta,Yuan:2014rca,Petrovic:2014xra,Brandt:2015ula,Hooper:2016rap}.\footnote{Young pulsars have also been considered within this context~\cite{OLeary:2015qpx,OLeary:2016cwz}, but are predicted to generate gamma-ray emission with a disk-like morphology, inconsistent with the observed excess.} Two recent analyses of Fermi data have elevated this possibility by presenting evidence in favor of an unresolved point source population in this region of the sky~\cite{Lee:2015fea,Bartels:2015aea}. It is not yet possible, however, to discern whether these studies have, in fact, identified signatures of a faint point source population, or have instead been systematically affected by variations in the gamma-ray flux associated with the small-scale structure of the diffuse gamma-ray background (see, for example, Ref.~\cite{Horiuchi:2016zwu}). 

Several arguments have been leveled against an MSP interpretation of the gamma-ray excess. The first argument is based on a comparison of the gamma-ray luminosity function between observed MSPs and any posited MSP population in the Inner Galaxy. 

If the luminosity function of both populations is similar, then any MSP explanation for the gamma-ray excess would be expected to result in many more gamma-ray point sources in the Inner Galaxy than have been reported~\cite{Hooper:2015jlu,Cholis:2014lta,SiegalGaskins:2010mp}. Second, no plausible mechanism has been identified to explain how so many gamma-ray bright MSPs might have come to populate the Inner Galaxy. Although tidally disrupted globular clusters are expected to deposit a population of MSPs into the Inner Galaxy~\cite{Brandt:2015ula,Gnedin:2013cda}, the gamma-ray emission from these sources is expected to generate only a few percent of the observed excess, once the effects of MSP evolution (\ie spin-down) are accounted for~\cite{Hooper:2016rap}. 

In order for MSPs from disrupted globular clusters to produce the total luminosity of the gamma-ray excess, the necessary mass of the central stellar cluster would have to drastically exceed observations~\cite{Hooper:2016rap}. Third, it was argued in Ref.~\cite{Cholis:2014lta} that such a population of MSPs in the Inner Galaxy should be accompanied by a much larger population of low-mass X-ray binaries (LMXBs) than has been observed. As LMXBs are known to be the progenitors of MSPs, it is difficult to imagine any scenario in which a sufficiently large number of MSPs could be present in the Inner Galaxy without being accompanied by a comparably large population of LMXBs.

In this paper, we revisit this connection between MSPs and LMXBs. Specifically, we study the gamma-ray emission from the Milky Way's globular cluster population, as well as the population of LMXBs and LMXB candidates that have been observed in these systems. Under the reasonable assumption that the number of LMXBs per unit gamma-ray emission from MSPs is similar in the Inner Galaxy to the ratio measured in globular clusters, we can calculate the flux of gamma rays expected from unresolved MSPs near the Galactic Center. We conduct this exercise and derive an upper limit on the fraction of the Galactic Center gamma-ray excess that originates from MSPs. We conclude that the relatively small number of bright LMXBs present in the Inner Galaxy disfavors MSP interpretations of Fermi's gamma-ray excess.

\section{Gamma-Ray Emission and LMXBs in Globular Clusters}
\label{clusters}

The gamma-ray emission observed from globular clusters~\cite{collaboration:2010bb,TheFermi-LAT:2015hja,Cholis:2014noa} is thought to originate from the MSPs contained within these systems~\cite{collaboration:2010bb,msps_47Tuc, msps_10_47Tuc, msps_more_47Tuc, Camilo:1999fc,Bogdanov:2006ap,Harding:2004hj,Bednarek:2007nn,Venter:2009mq,Cheng:2010za}. Due to the significant overabundance of MSPs in globular clusters, it is widely believed that these pulsars are generated through stellar interactions~\cite{1992xbfb.work..349J,2005ASPC..328..147C,2008MNRAS.386..553I,2010ApJ...714.1149H,2011ApJ...730L..11L,Bahramian:2013ihw,2013MNRAS.436.3720T}. For this reason, in this study we focus on those globular clusters with high stellar encounter rates, $\Gamma_e > 0.5$ (in units such that $\Gamma_e=1$ for NGC 104), as calculated in Ref.~\cite{Bahramian:2013ihw} (our results do not depend strongly on the precise choice of this cut). In Table~\ref{table2}, we list the 22 Milky Way globular clusters with $\Gamma_e > 0.5$~\cite{Bahramian:2013ihw}, as well as their distances~\cite{1996AJ....112.1487H} and gamma-ray fluxes as measured by the Fermi Gamma-Ray Space Telescope. Note that although we do not take into account distance uncertainties in our analysis, these uncertainties are negligible to our final results, which are dominated by other factors, such as the Poisson errors associated with the number of sources.

In analyzing the publicly available Fermi-LAT data, we have followed the same procedure as described in Ref.~\cite{Hooper:2016rap}, except here we have adopted a MSP-like spectral shape~\cite{Cholis:2014noa}: 
\begin{equation}
\frac{dN_{\gamma}}{dE_{\gamma}} \propto E_{\gamma}^{-1.57} \, \exp(-E_{\gamma}/3.78 \, {\rm GeV}).
\end{equation}
We have utilized 85 months of Pass 8 Source event class data in our analysis, applying standard cuts, and binning the data into 15 logarithmic energy bins between 0.1 and 100 GeV, as well as 280$\times$280 angular bins spanning a 14$^\circ\times$14$^\circ$ region-of-interest centered on the position of the source. By determining the maximum improvement to the likelihood, we calculate the value of the test statistic (TS) for each globular cluster, as well as the likelihood profiles for the gamma-ray flux. The errors quoted for the fluxes in Table~\ref{table2} denote the 1$\sigma$ range, as determined using the full likelihood profile. Stacking this list of sources, we find that the total gamma-ray luminosity from this collection of globular clusters is given by $L_{\gamma} = 1.39^{+0.04}_{-0.05}  \times 10^{36}$  erg$/$s, $>$0.1 GeV. After applying a mirrored sky location test~\cite{Hooper:2016rap,Linden:2015qha}, we confirm that our false detection rate is expected to be small, and that spurious detections are unlikely to significantly impact this determination.\footnote{We note that any spurious detections of globular clusters by Fermi, such as those resulting from the mismodeling of background fluctuations, would increase our estimate for the the total gamma-ray luminosity from MSPs in the Inner Galaxy, making the results presented here conservative.} For further information pertaining to our background model, including the treatment of point sources, we direct the reader to Ref.~\cite{Hooper:2016rap}.\footnote{The gamma-ray fluxes from the globular clusters NGC 6624 and NGC 6626 differ significantly from their values as presented in Ref.~\cite{Hooper:2016rap}. These two clusters each contain a source that is listed in the 3FGL catalog, whose spectrum was allowed to float in the previous analysis. Here, the fluxes shown reflect the total flux from each cluster, including that from any 3FGL sources that they may contain.}

\begin{table}
\renewcommand{\arraystretch}{1.2}
\begin{tabular}{|c|c|c|c|c|c|c|}
\hline 
Globular Cluster & Flux (erg/cm$^2$/s) & Distance (kpc) & Stellar Encounter Rate & TS \tabularnewline
\hline 
\hline 
NGC 104   &  $2.51^{+0.05}_{-0.06} \times 10^{-11}$ & 4.46 & 1.00 &  3995.9  \tabularnewline
 \hline
 NGC 362 & $6.74^{+2.63}_{-2.46} \times 10^{-13}$ &  8.61 &     0.74     & 9.69 \tabularnewline
 \hline
 Palomar 2 & $<2.69 \times 10^{-13}$ &    27.11 &   0.93  &  0.0  \tabularnewline
 \hline 
 NGC 6624 &   $1.14^{+0.10}_{-0.10} \times 10^{-11}$ & 7.91 & 1.15 & 455.8 \tabularnewline
 \hline
NGC 1851 &          $    9.05^{+ 2.92}_{- 2.67} \times 10^{-13}$ & 12.1 &  1.53 &   14.4 \tabularnewline
 \hline
 NGC 5824 & $<4.78 \times 10^{-13}$ & 32.17 &   0.98  &   0.0  \tabularnewline
 \hline 
 NGC 6093 & $4.32^{+0.57}_{-0.53} \times 10^{-12}$ &      10.01 &   0.53  & 91.9    \tabularnewline
 \hline 
 NGC 6266 & $1.84^{+0.07}_{-0.10} \times 10^{-11}$ &   6.83  &   1.67  &  850.7  \tabularnewline
 \hline  
  NGC 6284 & $<2.85 \times 10^{-13}$ &   15.29 &   0.67    & 0.0   \tabularnewline
 \hline   
NGC 6441            & $    1.00^{+ 0.09}_{- 0.07} \times 10^{-11}$& 11.6 &  2.30 &   210.9 \tabularnewline
 \hline
NGC 6652       & $    4.84^{+ 0.51}_{- 0.52} \times 10^{-12}$  & 10.0 &  0.70 &   128.3 \tabularnewline
 \hline
NGC 7078/M15  & $    1.81^{+ 0.40}_{- 0.39} \times 10^{-12}$& 10.4 & 4.51  &   29.7 \tabularnewline
 \hline
NGC 6440    & $    1.57^{+ 0.10}_{- 0.11} \times 10^{-11}$ & 8.45 &  1.40 &  311.2 \tabularnewline
 \hline
Terzan 6          &   $    2.18^{+ 1.20}_{- 0.90} \times 10^{-12}$ & 6.78 & 2.47 &   5.1 \tabularnewline
 \hline
NGC 6388               & $    1.77^{+ 0.06}_{-0.09} \times 10^{-11}$& 9.92 &   0.90 &   778.4 \tabularnewline
 \hline
NGC 6626/M28  & $1.95^{+0.13}_{-0.13} \times 10^{-11}$ & 5.52 &  0.65 & 749.8 \tabularnewline
 \hline
Terzan 5 &         $    6.61^{+ 0.17}_{- 0.13} \times 10^{-11}$ & 5.98 &6.80 &  2707.1 \tabularnewline
 \hline
NGC 6293 & $9.39^{+5.69}_{-5.45} \times 10^{-13}$ &      9.48    & 0.85  &   3.98 \tabularnewline
 \hline
NGC 6681 & $9.91^{+4.14}_{-3.86} \times 10^{-13}$  &      9.01 &    1.04   & 7.2 \tabularnewline
 \hline
 NGC 2808      & $    3.77^{+ 0.48}_{- 0.48} \times 10^{-11}$ & 9.59 &  0.92 &  96.7 \tabularnewline
 \hline
NGC 6715 & $6.02^{+4.15}_{-3.77} \times 10^{-13}$ &       26.49  &  2.52  & 2.6   \tabularnewline
 \hline
NGC 7089 &  $<4.50 \times 10^{-13}$  &                         11.56 &   0.52    & 0.0 \tabularnewline
 \hline 
  \end{tabular}
\caption{The gamma-ray fluxes (integrated between 0.1 and 100 GeV), distances~\cite{1996AJ....112.1487H}, and stellar encounter rates of the 22 globular clusters in the Milky Way with stellar encounter rates of $\Gamma_e > 0.5$, as calculated in Ref.~\cite{Bahramian:2013ihw} and in units such that the rate for NGC 104 is equal to unity. In calculating the gamma-ray fluxes and test statistic (TS), we have adopted a millisecond pulsar-like spectral shape, $dN_{\gamma}/dE_{\gamma} \propto E_{\gamma}^{-1.57} \exp(-E_{\gamma}/3.78 \, {\rm GeV})$.}
\label{table2}
\end{table}

\begin{table}
\renewcommand{\arraystretch}{1.2}
\begin{tabular}{|c|c|c|c|c|c|c|}
\hline 
LMXB &Notes & Globular Cluster & References \tabularnewline
\hline 
\hline 
4U 1820-30 &P&NGC 6624 & \cite{1974ApJS...27...37G,2007ApJ...654..494T,1987ApJ...312L..17S} \tabularnewline
 \hline
4U 0513-40 &P&NGC 1851  & \cite{1975ApJ...199L.143C,2011MNRAS.414L..41F,2009ApJ...699.1113Z} \tabularnewline
 \hline
4U 1746-37 &P&    NGC 6441 &   \cite{1974ApJS...27...37G,2004ApJ...607L..33B,BalucinskaChurch:2003dz}        \tabularnewline
 \hline
XB 1832-330 &P&  NGC 6652 &  \cite{2004ApJ...607L..33B,1985ApJ...290..171H,Engel:2012uw}     \tabularnewline
 \hline
M15 X-2 &P&   NGC 7078/M15 &  \cite{2001ApJ...561L.101W,2007AA...475..775K,Dieball:2005px} \tabularnewline
 \hline
AC 211 & P & NGC 7078/M15 &  \cite{1974ApJS...27...37G,2007AA...475..775K,1993AA...270..139I}  \tabularnewline
 \hline
SAX J1748.9-2021 &T, XP& NGC 6440 &  \cite{1975Natur.257...32M,2004ApJ...607L..33B,Altamirano:2007ru}  \tabularnewline
 \hline
GRS 1747-312 &T&  Terzan 6    &  \cite{1991AA...246L..40P,2006ATel..734....1C,in'tZand:2003eq}     \tabularnewline
 \hline
Terzan 6 X-2 &T&  Terzan 6 & \cite{2016PASJ...68S..15S} \tabularnewline
 \hline
IGR J17361-4441 &T&   NGC 6388 &  \cite{2011ATel.3565....1G,Bozzo:2011qt}              \tabularnewline
 \hline
IGR J18245-2542 &T, XP&  NGC 6626/M28 & \cite{2013Natur.501..517P,Papitto:2012su}  \tabularnewline
 \hline
EXO 1745-248 &T& Terzan 5  & \cite{1981ApJ...247L..23M,Galloway:2006eq}   \tabularnewline
 \hline
IGR J17480-2446 &T&Terzan 5 & \cite{2010ATel.2919....1B,2010ATel.2946....1S,Papitto:2010bu} \tabularnewline
 \hline
Terzan 5 X-3 &T&Terzan 5  & \cite{2014ApJ...780..127B} \tabularnewline
 \hline 
MAXI J0911-635 &T&NGC 2808 & \cite{2016arXiv161102995S}    \tabularnewline
 \hline
  \end{tabular}
\caption{The list of LMXBs in globular clusters with stellar encounter rates $\Gamma_e>0.5$ observed between 2003 and 2016 that would have been detected by INTEGRAL if they had been located in the Inner Galaxy. In particular, these sources each reached an X-ray luminosity of $\gsim 10^{36}$ erg$/$s for a duration of a week or more. The codes in the ``Notes'' column denote whether a given LMXB is a persistent source (P), transient source (T), and/or an X-ray pulsar (XP). For each LMXB, the references listed include the detection by INTEGRAL.}
\label{table1}
\end{table}

The INTEGRAL telescope provides us with our most sensitive and complete catalog of LMXBs in the Inner Galaxy, and we make use of these observations in Sec.~\ref{ig} to characterize the LMXB population in and around the Galactic Center. In order to facilitate a comparison between the LMXBs found within globular clusters and those found within the Inner Galaxy, we have compiled a list of those LMXBs in our sample of 22 globular clusters (those with $\Gamma_e >0.5$) that would almost certainly have been detected by INTEGRAL if they had instead been located in the Inner Galaxy (see Table~\ref{table1}). More specifically, we consider a given LMXB to be detectable by INTEGRAL (if it had been located in the Inner Galaxy) if it has reached an X-ray luminosity exceeding $10^{36}$ erg$/$s for a duration of at least a week at some point over the period of INTEGRAL's mission. This is a conservative choice of threshold, as INTEGRAL has collected sufficient exposure from the region around the Galactic Center to detect significantly fainter sources if they have been active for a long time. To compile this list, we begin with the 18 (non-quiescent) sources listed in Table 5 of Ref.~\cite{2014ApJ...780..127B}, which contains all such LMXBs found in globular clusters, published as of 2014. To make a fair comparison with the collection of sources detected by INTEGRAL, we remove NGC 6440 X-2 from this list, as its peak luminosity of $L_{X} \sim (2-3) \times 10^{36}$ erg/s was only reached over a timescale of a day, and it wasn't detected by INTEGRAL, or even clearly detected by the Rossi X-ray Timing Explorer (RXTE/PCA) bulge monitoring, which is more sensitive~\cite{2010ApJ...714..894H}. We also removed the source XB 1732-304 from our list, as it has been quiescent since 1999, and thus does not overlap with the period covered by INTEGRAL. The other 16 LMXBs listed in Ref.~\cite{2014ApJ...780..127B} each have peak X-ray luminosities and durations large enough to have been detected by INTEGRAL if they had been located in the Inner Galaxy. In fact, we note that many of these LMXBs were originally detected by INTEGRAL (as denoted by the IGR names in Table~\ref{table1}). We then add to this list two more recently discovered transient sources which reached a luminosity of $\sim$\,$10^{36}$  erg/s in 2016 (in NGC 2808) and in 2009 (in Terzan 6), respectively~\cite{Sanna:2016xmm,2016PASJ...68S..15S}. We have removed the sources X1850-087 (in NGC 6712), 4U 1722-30 (in Terzan 2) and MXB 1730-335 (in Liller 1) from our sample, as $\Gamma_e < 0.5$ for each of these clusters. There are 15 globular cluster LMXBs  that meet our selection criteria, shown in Table~\ref{table1}.

As a consequence of source confusion, it is likely that observations have missed some transient X-ray binary outbursts from globular clusters which were masked by the presence of a bright persistent LMXB. For example, Galloway {\it et al}.~\cite{Galloway:2006eq} presented evidence from the patterns of thermonuclear bursts in NGC 6441 that more than one LMXB was active in the cluster at two separate times (both earlier than our considered date range). To estimate how many X-ray binary outbursts were likely missed, we calculated the number of transient LMXB outbursts seen in globular clusters with and without persistent LMXBs (within our considered timeframe). In globular clusters without persistent LMXBs, we found 9 transient systems. The total encounter rate of these clusters is 24.6 (in units where the encounter rate of NGC 104 is unity), corresponding to a rate of $0.37^{+0.16}_{-0.13}$ transients per unit encounter rate. We then applied this to the five clusters which host persistent sources, which sum to a total normalized encounter rate of 10.2 (we neglect errors associated with encounter rates, which do not substantially increase the final uncertainties). From this exercise, we estimate that $3.7^{+2.9}_{-2.7}$ transients are likely to be missing from the list shown in Table~\ref{table1} (including Poisson errors).

\section{LMXBs in the Inner Galaxy}
\label{ig}

To establish and constrain the population of LMXBs in the Inner Galaxy, we make use of the INTEGRAL General Reference Catalog~\cite{IGSC}, which was constructed by merging several catalogs available in the HEASARC database~\cite{2000A&AS..147...25L,2001A&A...368.1021L,1999ApJS..120..335M,1978ApJS...38..357F,1984ApJS...54..581L,1999ApJ...519..637M,2001AIPC..599..991T,2001ApJS..134...77S,2002ApJS..138...19S}. From this catalog, we consider all sources within a $10^{\circ}$ radius around the Galactic Center that were detected by either (or both) ISGRI or JEM-X, and that are listed as either an LMXB or as an unclassified source. We remove those sources that have been categorized as a supergiant fast X-ray transient, high-mass X-ray binary, black hole candidate, symbiotic X-ray binary, or as spurious. After applying these criteria, we are left with a collection of 42 LMXBs, along with 46 unclassified sources.

\begin{figure}
\includegraphics[keepaspectratio,width=0.412\textwidth]{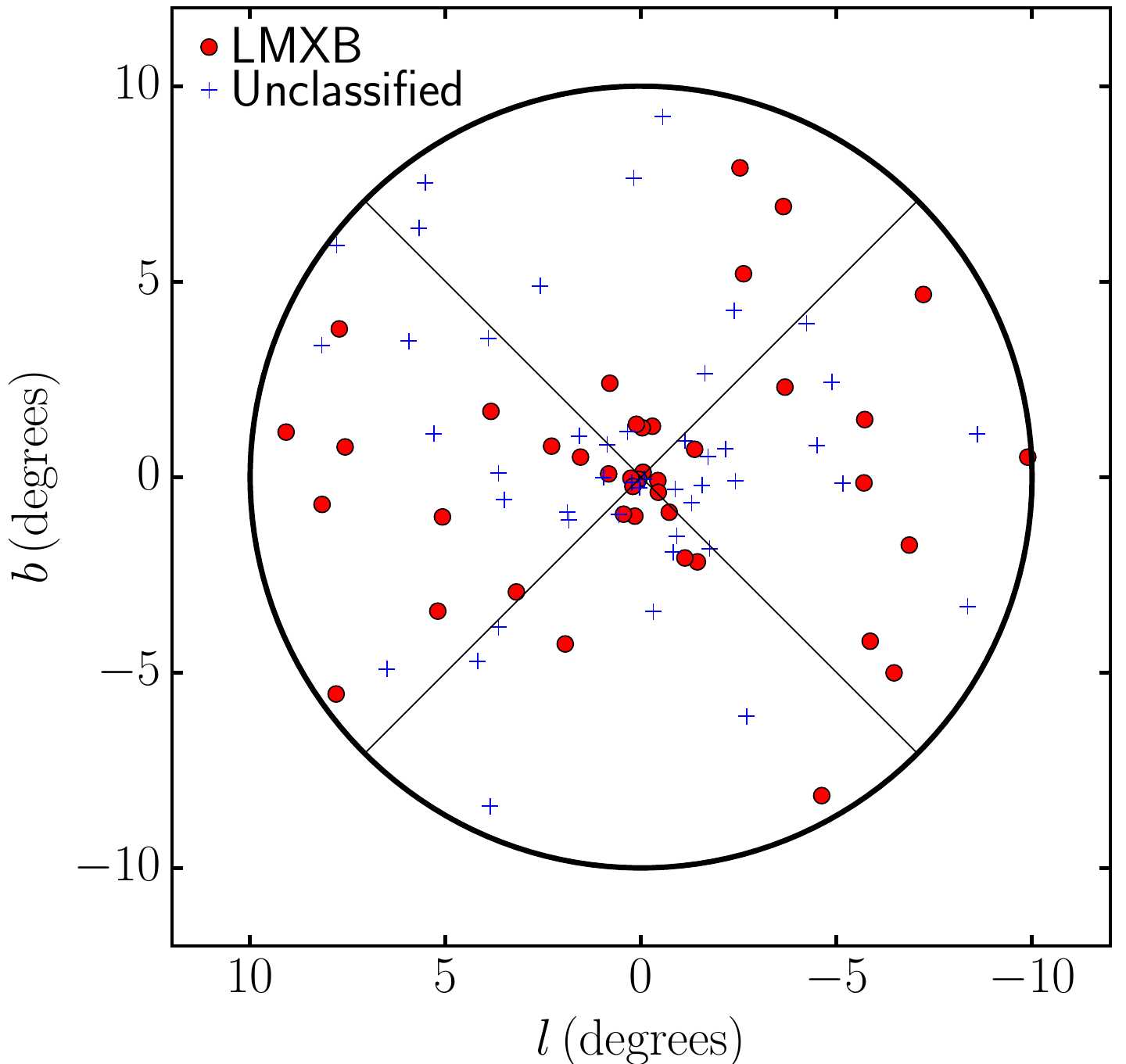}
\hspace{0.2cm}
\includegraphics[keepaspectratio,width=0.495\textwidth]{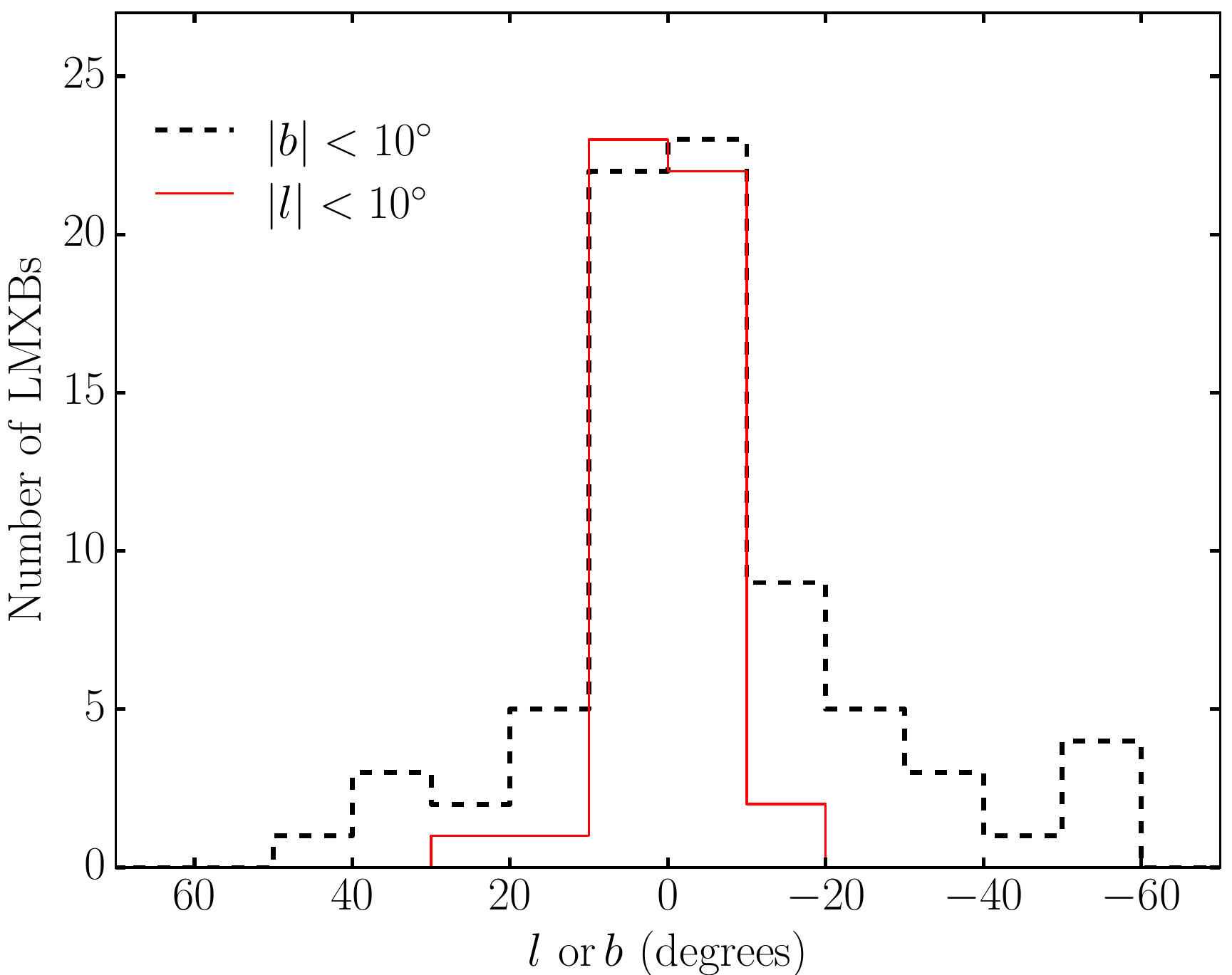}
\caption{Left: The distribution of X-ray sources characterized as an LMXB (red circles) or as an unclassified source (blue crosses) within the innermost ten degrees around the Galactic Center, as given in the INTEGRAL Reference Source Catalog. We do not include those sources categorized as a supergiant fast X-ray transient, high-mass X-ray binary, black hole candidate, symbiotic X-ray binary, or as spurious. Right: The distribution of sources characterized as an LMXB as a function of galactic longitude (latitude) for sources that reside within $|b|<10^{\circ}$ ($|l|<10^{\circ}$).}
\label{lmxbmap}
\end{figure}

Of course, not all LMXBs observed in the innermost 10$^{\circ}$ around the Galactic Center are members of a population associated with the Inner Galaxy. Some are instead part of a population associated with the Galactic Disk. In Fig.~\ref{lmxbmap}, we present the distribution of these sources as observed on the sky. In the left frame, we plot the distribution of those sources characterized as LMXBs (red circles) or as unclassified (blue crosses) within the innermost ten degrees around the Galactic Center. We further divide this region into four subregions of equal solid area, finding that 17 (18) of the sources classified as LMXBs (unclassified) are present in the north and south subregions, while 25 (28) are present in the west and east subregions. This asymmetry between the number of sources in the north+south and west+east subregions provides evidence for a disk population of sources, in addition to any bulge or Inner Galaxy population that may be present. We further explore the morphology of observed LMXBs in the right frame of this figure, where we plot the distribution of those sources characterized as an LMXB as a function of galactic longitude (latitude) for sources that reside within $|b|<10^{\circ}$ ($|l|<10^{\circ}$). This demonstrates that there are more sources distributed along the plane than either north or south of the Inner Galaxy, again supporting the conclusion that a significant fraction of the observed LMXBs are not part of a bulge or Inner Galaxy population.

To quantify and constrain the fraction of LMXBs in the direction of the Inner Galaxy that are either part of a spherically distributed population or a population associated with the Galactic Disk, we construct a simple morphological model. We begin with a spherical population, with a number density of sources that is described as follows:  
\begin{equation}
n(r) \propto r^{-\Gamma},
\label{profile}
\end{equation}
where $r$ is the distance to the Galactic Center. In order for such a source population to generate a gamma-ray morphology that is consistent with the Galactic Center excess, we require $\Gamma \simeq 2.0$ to 2.6 (corresponding to a dark matter density profile with an inner slope of $\gamma \simeq 1.0$ to 1.3). The projected distribution of this source population on the sky is given by:
\begin{equation}
P_{\rm sph}(\psi) \propto \int_{los} n(r) dl,
\end{equation}
where $\psi$ is the angle observed and the integral is performed over the line-of-sight (los). In addition to this spherical source population, we further postulate a population associated with the Galactic Disk, with a projected distribution described as follows:
\begin{equation}
P_{\rm disk}(l,b) \propto \exp\bigg(-\frac{|b|}{b_{\rm scale}}\bigg),
\end{equation}
where we treat the scale height, $b_{\rm scale}$, as a free parameter. Combining these two populations, we write:
\begin{equation}
P(l,b) = (1-f_{\rm disk}) \,  P_{\rm sph} (l,b)  +  f_{\rm disk} \,  P_{\rm disk} (l,b), 
\label{sum}
\end{equation}
where $P_{\rm sph}$ and $P_{\rm disk}$ are each individually normalized to the number of sources observed within the innermost $10^{\circ}$ around the Galactic Center (see the left frame of Fig.~\ref{lmxbmap}) and $f_{\rm disk}$ is the fraction of those sources that are part of the disk population.

\begin{figure}
\includegraphics[keepaspectratio,width=0.6\textwidth]{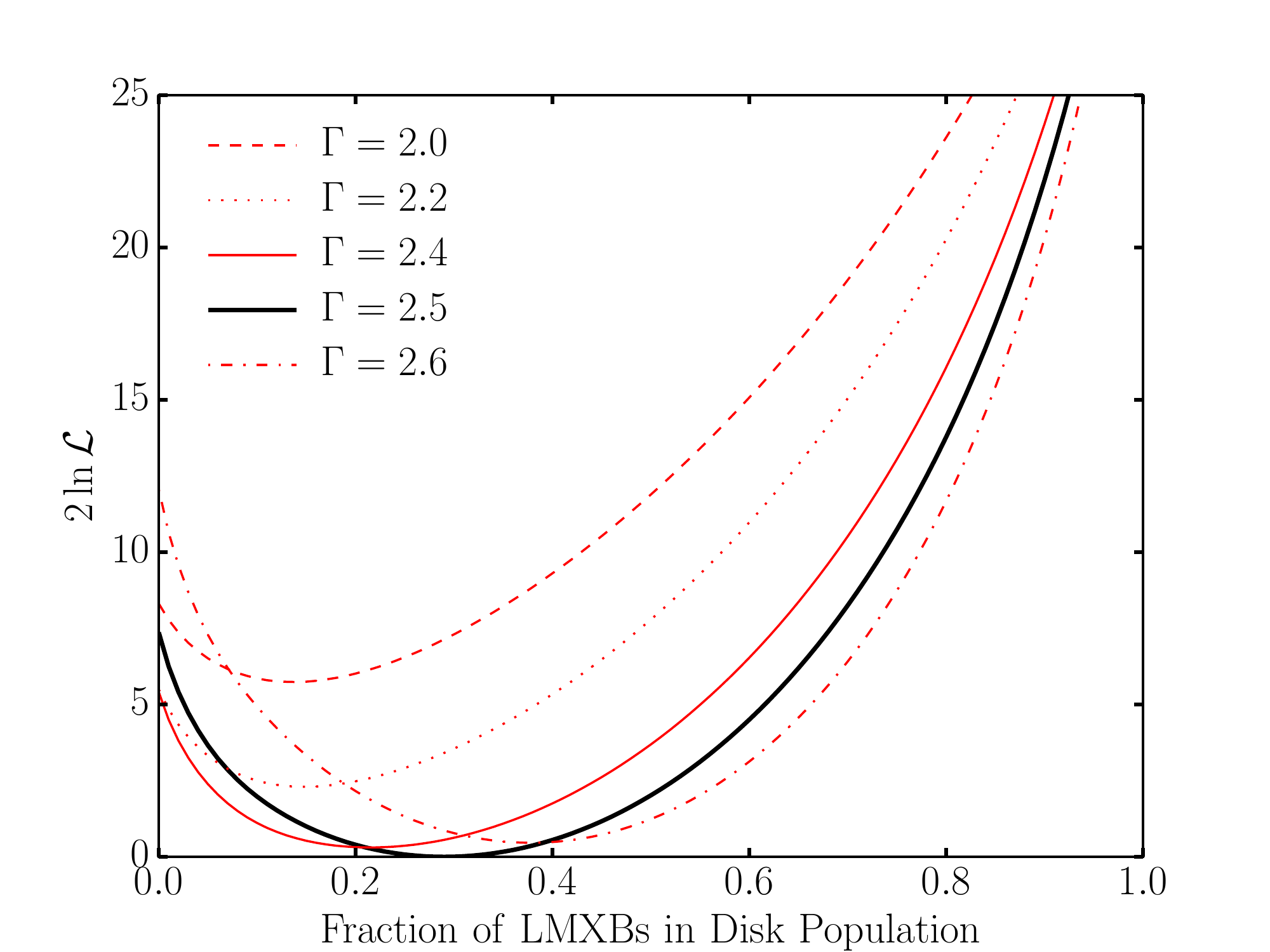}
\caption{Using the distribution of LMXBs shown in the left frame of Fig.~\ref{lmxbmap}, we have fit the parameters of a model consisting of spherical and disk populations (see Eqns~\ref{profile}-\ref{sum}). After marginalizing over the width of the disk distribution, $b_{\rm scale}$, we plot twice the log-likelihood of the fit as a function of the fraction of the LMXBs that are part of the disk population. We show results for several values of the slope of the density profile of the spherical component, $n(r) \propto r^{-\Gamma}$. After marginalizing over both $\Gamma$ and $b_{\rm scale}$, we conclude that $33^{+15}_{-22}\%$ of the Inner Galaxy LMXBs belong to a disk-like population.}
\label{diskfrac}
\end{figure}

For a range of values of $b_{\rm scale}$, $f_{\rm disk}$, and $\Gamma$, we have calculated the predicted angular distribution of LMXBs and compared this to the observed distribution (discretizing the data into $1^{\circ} \times 1^{\circ}$ bins). We find a best-fit for $\Gamma=2.5$, $b_{\rm scale}=2.4^{\circ}$, and $F_{\rm disk}=0.33$, with variations as shown in Fig.~\ref{diskfrac}. After marginalizing over both $\Gamma$ and $b_{\rm scale}$, we find that the fit to the observed distribution yields the following fraction of sources associated with a disk population: $F_{\rm disk} = 0.33^{+0.15}_{-0.22}$.

Note that in deriving this result, we have not relied in any way on the distribution of sources outside of the innermost $10^{\circ}$. That being said, the distribution of sources shown in the right frame of Fig.~\ref{lmxbmap} also supports the conclusion that roughly a third of the LMXBs in the innermost $10^{\circ}$ are members of a disk-like population. 

In this calculation, we have neglected variations in INTEGRAL's exposure over the solid angle of the Inner Galaxy.  Although INTEGRAL's sensitivity to LMXBs is approximately radially symmetric over this region, its exposure falls somewhat with angular distance from the Galactic Center. More specifically, INTEGRAL's exposure is approximately flat within the central $5^{\circ}$--\,$6^{\circ}$, but then drops below 50\% of the maximum at about $9^{\circ}$ from the Galactic Center (see, for example, Fig.~1 of Ref.~\cite{Bird:2016rai}). This could subtly affect the parameters of the inferred LMXB distribution. In particular, the true value of slope, $\Gamma$, is likely to be slightly smaller (flatter) than shown in Fig.~\ref{diskfrac}. We also note that source confusion within the innermost $\sim$\,$0.5^{\circ}$ around the Galactic Center could slightly impact the distribution of LMXBs that is favored by our fit. 

To estimate the impact of the non-uniformity of INTEGRAL's exposure, we repeated out fit calculation adopting a model in which INTEGRAL's sensitivity drops by a factor of 50\% between $5^{\circ}$ and $10^{\circ}$ from the Galactic Center, and adopting the LMXB luminosity function described in Ref.~\cite{2007AA...475..775K}. This had the net effect of increasing the fraction of LMXBs that are associated with the disk population from 33\% to 49\% in our best-fit model, while negligibly impacting the preferred value of $\Gamma$.

\section{Constraints on MSPs in the Inner Galaxy}
 
From the information presented in the previous two sections, we can calculate the gamma-ray flux that is predicted from MSPs in the inner region of the Milky Way.  In particular, these quantities are related as follows:
\begin{eqnarray}
L_{\gamma}^{\rm IG} &=& L_{\gamma}^{\rm clusters} \times \bigg(\frac{N^{\rm IG}_{\rm LMXB}}{N_{\rm LMXB}}\bigg),\label{maineq}
\end{eqnarray}
where $L_{\gamma}^{\rm clusters}$  is the gamma-ray luminosity from the stacked sample of 22 globular clusters listed in Table~\ref{table2} and $N_{\rm LMXB}$ is the number of (INTEGRAL detectable) LMXBs present in that same collection of globular clusters. We compare this to $N^{\rm IG}_{\rm LMXB}$, which is the number of LMXBs detected by INTEGRAL in the Inner Galaxy. We remind the reader that this relationship is predicated on the assumption that the relative populations of LMXBs and MSPs are the same in globular clusters and in the Inner Galaxy. We revisit and discuss this assumption further in Sec.~\ref{five}.

We start by making a very conservative (and unrealistic) assumption that none of the MSPs in our sample of globular clusters would be resolved by Fermi if they were instead located in the Inner Galaxy. In this case, $L^{\rm clusters}_{\gamma}$ is set to the full value as determined in Sec.~\ref{clusters}, $L^{\rm clusters}_{\gamma} = 1.39^{+0.04}_{-0.05}  \times 10^{36}$ erg$/$s. For $N^{\rm IG}_{\rm LMXB}$, we adopt values between $42 \times (1-F_{\rm disk})$ and $88 \times (1-F_{\rm disk})$, depending on the fraction of unclassified INTEGRAL sources in the Inner Galaxy region that are in fact LMXBs. Lastly, we take $N_{\rm LMXB}=18.7^{+4.7}_{-4.5}$, corresponding to the list of sources given in Table~\ref{table1} and including Poisson errors as well as our estimate for the number of undetected transients. Combining the errors on the fitted value of $F_{\rm disk}$ with the poissonian errors on the observed numbers of observed LMXBs, we arrive at the following determination:

\begin{eqnarray}
L_{\gamma}^{\rm IG} &=& (2.09^{+0.86}_{-0.71}) \times 10^{36} \, {\rm erg/s}, \,\,\,\,\,\,\,\,\, {\rm Only}\,\, {\rm Sources}\,\, {\rm Classified}\,\, {\rm as} \,\,{\rm LMXBs} \nonumber \\
L_{\gamma}^{\rm IG} &=& (4.38^{+1.79}_{-1.48}) \times 10^{36} \, {\rm erg/s}, \,\,\,\,\,\,\,\,\, {\rm Including}\,\,{\rm All} \,\, {\rm Unclassified}\,\, {\rm Sources} 
\label{maineq1}
\end{eqnarray}
Comparing this result with the measured gamma-ray luminosity of gamma-ray excess, \,\,\, $L_{\gamma} = (2.0 \pm 0.4) \times 10^{37}$ erg$/$s integrated within $10^{\circ}$ of the Galactic Center~\cite{Calore:2014xka,cholisPC}, we estimate that $10.5^{+4.7}_{-4.1}\%$ (only LMXBs) or $21.9^{+9.9}_{-8.6}\%$ (LMXBs and unclassified) of the excess emission can be potentially attributed to an underlying MSP population. As mentioned above, however, this calculation almost certainly overestimates the fraction of the Galactic Center excess that arises from MSPs. 

As the emission from gamma-ray bright globular clusters is in many cases dominated by the brightest individual MSP in that system~\cite{Hooper:2016rap,2011Sci...334.1107F,TheFermi-LAT:2013ssa,Johnson:2013uza} (measured pulsations, in fact, confirm that the gamma-ray emission from NGC 6624 and NGC 6626 are dominated by the individual pulsars PSR J1823-3021A~\cite{2011Sci...334.1107F,TheFermi-LAT:2013ssa} and PSR B1821-24~\cite{Johnson:2013uza}, respectively), we expect that a significant fraction of the gamma-ray emission from these MSPs would be resolved as point sources, and thus would not contribute to the Galactic Center excess (sources in the Fermi Third Source Catalog are generally included in the background models of such analyses). The gamma-ray emission from several of the globular clusters considered in this study (listed in Table~\ref{table2}) is very bright, leading these sources to easily be resolved as individual sources by Fermi, even if they had been located in the Inner Galaxy. In fact, the globular clusters NGC 6441, NGC 6440 and Terzan 5 are located within the innermost $10^{\circ}$ around the Galactic Center and are contained within the 3FGL source catalog. In addition, the globular clusters NGC 6266, NGC 6388 or NGC 2808 (each with $L_{\gamma} >10^{35}$ erg/s) are well above Fermi's detection threshold and would be very likely to have been detected by Fermi and contained within the 3FGL if located in the Inner Galaxy. To quantify how much of the gamma-ray emission from such globular clusters is likely to originate from MSPs that are detectable as individual point sources by Fermi, we utilize a Monte Carlo to draw from the MSP luminosity function as determined in Ref.~\cite{Hooper:2016rap}, finding that from among these six globular clusters (NGC 6441, NGC 6440, Terzan 5, NGC 6266, NGC 6388 and NGC 2808), $41.9^{+10.3}_{-12.9}\%$ of their gamma-ray emission comes from MSPs with individual luminosities greater than $5 \times 10^{34}$ erg/s, which we take as Fermi's approximate point source threshold in this region (the approximate flux of the faintest Inner Galaxy sources contained in the 3FGL catalog). In light of this, we remove the resolved component from the stacked gamma-ray flux, which has the impact of decreasing the predicted gamma-ray flux from unresolved MSPs in the Inner Galaxy from that presented in Eq.~\ref{maineq1} to the following, more realistic, estimate:
\begin{eqnarray}
L_{\gamma}^{\rm IG} &=& (1.50^{+0.65}_{-0.54}) \times 10^{36} \, {\rm erg/s}, \,\,\,\,\,\,\,\,\, {\rm Only}\,\, {\rm Sources}\,\, {\rm Classified}\,\, {\rm as} \,\,{\rm LMXBs} \nonumber \\
L_{\gamma}^{\rm IG} &=& (3.15^{+1.37}_{-1.12}) \times 10^{36} \, {\rm erg/s}, \,\,\,\,\,\,\,\,\, {\rm Including}\,\,{\rm All}\,\, {\rm Unclassified}\,\, {\rm Sources}, 
\label{maineq2}
\end{eqnarray}
which corresponds to $7.5^{+3.6}_{-3.1}\%$ (only LMXBs) or $15.7^{+7.5}_{-6.4}\%$ (LMXBs and unclassified) of the emission associated with the Galactic Center excess. We further conclude that no more than 28.3\% (95\% CL, conservatively including all unclassified sources) of the Galactic Center gamma-ray excess originates from a centrally located population of MSPs, assuming that the luminosity function of MSPs and their abundance relative to LMXBs is similar to that observed in globular clusters. Put another way, we predict that a gamma-ray flux from MSPs that is as large as the observed Galactic Center excess should be accompanied by approximately $\sim 10^3$ detectable LMXBs; far more than the 42 (88) LMXBs (LMXBs and LMXB candidates) that have actually been observed by INTEGRAL. We note that this result is consistent with that presented previously in Ref.~\cite{Cholis:2014lta}, but is more robust and has much smaller error bars due to the larger number of sources included in this analysis.

\section{Luminosity Function Evolution?}
\label{five}

The hypothesis that the Galactic Center gamma-ray excess could be generated by a population of unresolved MSPs is in considerable tension with both the number of gamma-ray sources and the number of LMXBs that have been observed in the direction of the Inner Galaxy.  This conclusion, however, relies on the assumption that the luminosity function of MSPs, and their abundance relative to LMXBs, is similar to that observed in globular clusters. In this section, we revisit this assumption and consider whether one might be able to imagine an MSP population that is capable of generating the observed gamma-ray excess without conflicting with these related observations. 

The population of LMXBs found within globular clusters is expected to be in an approximately steady-state configuration, with new LMXBs being formed through stellar encounters at a rate similar to those transitioning into an MSP phase. We would also expect this to be the case for the LMXBs and MSPs in the Inner Galaxy if they are produced through the evolution of the bulge's stellar population. If the Inner Galaxy's MSPs and LMXBs originate in tidally disrupted globular clusters, however, there could be a significant departure from steady-state. In particular, once a given globular cluster has migrated through dynamical friction into the Inner Galaxy and is tidally disrupted~\cite{Gnedin:2013cda,2012ApJ...750..111A,2013ApJ...763...62A,Bednarek:2013oha,Brandt:2015ula}, new LMXBs will largely cease to be formed, leading to the evolution of the combined LMXB/MSP population. Due to low stellar encounter rates, we do not expect significant numbers of new LMXBs to form in the Inner Galaxy.

After the formation of an LMXB progenitor (either as a primordial binary, or through a stellar encounter) such a system will remain intact for some time, often a Gyr, before making contact and beginning the process of mass transfer.\footnote{It is possible that some MSPs are formed from intermediate-mass X-ray binaries~\cite{2000ApJ...532L..47R,2002ApJ...565.1107P} or other non-LMXB progenitors. What is important to our analysis is that the fraction of MSPs that are formed from LMXBs is similar in the Inner Galaxy as is found in globular clusters.} Following this phase, a typical LMXB is thought to remain active for a period of time on the order of a Gyr. For example, Ref.~\cite{2013ApJ...764...41F} calculates that the X-ray emissivity of a given LMXB remains constant for about 1.5 Gyrs after formation, and then declines with an e-folding timescale of roughly 1.5 Gyrs (see also, Ref.~\cite{Humphrey:2006et}). From this, we expect the MSP-to-LMXB ratio associated with a given globular cluster to begin undergoing significant evolution only after a Gyr or more has passed since its tidal disruption. Thus, if the MSPs responsible for the gamma-ray excess were almost entirely deposited in the Inner Galaxy more than $\sim$\,1-3 Gyr ago, this could potentially explain the lack of observed LMXBs. 

Over the same period of time, however, the gamma-ray luminosity function of the MSP population will also evolve. The gamma-ray luminosity of an individual MSP is given by:
\begin{eqnarray}
\label{lum}
L_{\gamma} &=& \eta_{\gamma} \, \dot{E} \\
&=& \eta_{\gamma} \, \frac{4 \pi^2 I \dot{P}}{P^3} \nonumber \\
&\simeq & 9.6 \times 10^{33} \, {\rm erg/s} \,\, \bigg(\frac{\eta_{\gamma}}{0.2}\bigg) \,  \bigg(\frac{B}{10^{8.5}\, {\rm G}}\bigg)^2 \, \bigg(\frac{3 \, {\rm ms}}{P}\bigg)^4, \nonumber
\end{eqnarray}
where $\eta_{\gamma}$ is the gamma-ray efficiency, $\dot{E}$ is the spin-down rate, $I$ is the neutron star's moment of inertia and $P$ and $\dot{P}$ are the rotational period of the pulsar and its time derivative. The decline in the pulsar's rotational frequency results from magnetic-dipole braking, and is related to its magnetic field and period as follows:
\begin{equation}
\dot{P} \simeq 3.3 \times 10^{-20} \, \bigg(\frac{B}{10^{8.5}\, {\rm G}}\bigg)^2 \, \bigg(\frac{3 \, {\rm ms}}{P}\bigg),
\label{pdot}
\end{equation}
where $B$ is the strength of the magnetic field, and we have taken $I=10^{45}$ g cm$^2$. We can use this information to write the time-dependent gamma-ray luminosity for a given MSP:
\begin{eqnarray}
L_{\gamma}(t) = 9.6 \times 10^{33} \, {\rm erg/s} \, \times \bigg(\frac{\eta_{\gamma}}{0.2}\bigg) \bigg(\frac{B}{10^{8.5}{\rm G}}\bigg)^2 \Bigg[\bigg(\frac{B}{10^{8.5}{\rm G}}\bigg)^2 \bigg(\frac{t}{1.45 \,{\rm Gyr}}\bigg)+\bigg(\frac{P_0}{3 \, {\rm ms}}\bigg)^2    \Bigg]^{-2}, \nonumber \\
\end{eqnarray}
where $P_0$ is the initial period of the pulsar. 

After the debris of a collection of tidally disrupted globular clusters has evolved for several Gyrs or more, many of the LMXBs may have declined in luminosity, allowing them to go undetected by INTEGRAL. Over a similar timescale, however, most of the MSPs with a gamma-ray luminosity of $\sim$\,$10^{34}$ erg/s or more will also become significantly less bright. And while this may enable such MSPs to evade detection by Fermi, this evolution also has the effect of significantly decreasing the total gamma-ray flux from this population, thus requiring an even larger number of MSPs in order to generate the observed Galactic Center excess. For an initial MSP luminosity function similar to that measured in globular clusters, for example, the total gamma-ray luminosity of the MSP population is predicted to fall by an order of magnitude over a period of $\sim$\,1.4 Gyr~\cite{Hooper:2016rap}, requiring a much larger number of pulsars to generate the observed intensity of the gamma-ray excess. If we consider the model of Gnedin {\it et al.}~\cite{Gnedin:2013cda} (and as employed in Ref.~\cite{Brandt:2015ula}) to calculate the rate of globular cluster infall and tidal disruption, and adopt the best-fit MSP luminosity function of Ref.~\cite{Hooper:2016rap}, we indeed find that spin-down evolution reduces the total gamma-ray luminosity from the MSP population by a factor of between $\sim$\,5-15, depending on the assumed lifetime of the LMXB phase. Even if we consider the maximum possible quantity of disrupted globular clusters (such that they are responsible for the entire mass of the Millky Way's central stellar cluster), the accompanying MSPs are unlikely to generate more than a few percent of the observed gamma-ray excess~\cite{Hooper:2016rap}, consistent with the findings of this study.

\section{Summary and Conclusions}

In this paper, we have revisited the connection between millisecond pulsars (MSPs) and low-mass X-ray binaries (LMXBs) in an effort to estimate and constrain the gamma-ray emission from the population of unresolved MSPs in the inner volume of the Milky Way. Toward this end, we have utilized the populations of LMXBs observed in globular clusters and in the Inner Galaxy, as well as the gamma-ray emission measured from globular clusters. From this comparison, we estimate that only 
$7.5^{+3.6}_{-3.1}\%$ of the gamma-ray emission associated with the Galactic Center excess originates from MSPs. If we more conservatively assume that all of the unclassified LMXB candidates in the Inner Galaxy are in fact LMXBs, then our estimate increases to $15.7^{+7.5}_{-6.4}\%$. The relatively small number of LMXBs detected in the Inner Galaxy disfavors the possibility that the gamma-ray excess observed by Fermi originates primarily from a population of unresolved MSPs.

\bigskip
\bigskip
\bigskip

\textbf{Acknowledgments.} We would like to thank Ilias Cholis for helpful discussions. D.~Haggard acknowledges support from a Natural Sciences and Engineering Research Council of Canada (NSERC) Discovery Grant and a Fonds de recherche du Qu{\'e}bec -- Nature et Technologies Nouveaux Chercheurs Grant. C.~Heinke is supported by a Natural Sciences and Engineering Research Council of Canada (NSERC) Discovery Grant and a Discovery Accelerator Supplement Grant. D.~Hooper is supported by the US Department of Energy under contract DE-FG02-13ER41958. Fermilab is operated by Fermi Research Alliance, LLC, under Contract No. DE-AC02-07CH11359 with the US Department of Energy. T.~Linden acknowledges support from NSF Grant PHY-1404311 to John Beacom. We acknowledge the Ohio Supercomputer Center for providing support for this work.

\bibliography{lmxb.bib}
\bibliographystyle{JHEP}

\end{document}